\documentclass{elsart}
\pdfoutput=1

\usepackage[square,comma]{natbib}
\usepackage{graphicx}
\usepackage{pxfonts}
\usepackage{amssymb}
\usepackage{footnote}%!!!
\usepackage{multirow}
\usepackage{varwidth}
\usepackage{array}

\journal{}

\begin{document}

\thispagestyle{empty}
\begin{Large}
\textbf{DEUTSCHES ELEKTRONEN-SYNCHROTRON}

\textbf{\large{Ein Forschungszentrum der Helmholtz-Gemeinschaft}\\}
\end{Large}

DESY 14-047

April 2014

\begin{eqnarray}
\nonumber
\end{eqnarray}
\begin{center}
\begin{Large}
\textbf{The Full Potential of the Baseline SASE Undulators of the European XFEL}
\end{Large}
\begin{eqnarray}
\nonumber &&\cr \nonumber
\end{eqnarray}

\begin{large}
Ilya Agapov$^a$, Gianluca Geloni$^a$, Guangyao Feng$^b$, Vitali Kocharyan$^b$, Evgeni Saldin$^b$, Svitozar Serkez$^b$, Igor Zagorodnov$^b$
\end{large}

\textsl{\\$^a$European XFEL GmbH, Hamburg}
\begin{large}

\end{large}
\textsl{\\$^b$Deutsches Elektronen-Synchrotron DESY, Hamburg}

\begin{large}

\end{large}

\begin{eqnarray}
\nonumber
\end{eqnarray}
\begin{eqnarray}
\nonumber
\end{eqnarray}
ISSN 0418-9833
\begin{eqnarray}
\nonumber
\end{eqnarray}
\begin{large}
\textbf{NOTKESTRASSE 85 - 22607 HAMBURG}
\end{large}
\end{center}
%\end{widetext}
\clearpage
\newpage
\begin{frontmatter}



\title{The Full Potential of the Baseline SASE Undulators of the European XFEL}


\author[XFEL]{Ilya Agapov \thanksref{corr},}
\thanks[corr]{Corresponding Author. E-mail address: ilya.agapov@xfel.eu}
\author[XFEL]{Gianluca Geloni,}
\author[DESY]{Guangyao Feng,}
\author[DESY]{Vitali Kocharyan,}
\author[DESY]{Evgeni Saldin,}
\author[DESY]{Svitozar Serkez,}
\author[DESY]{Igor Zagorodnov}

\address[XFEL]{European XFEL GmbH, Hamburg, Germany}
\address[DESY]{Deutsches Elektronen-Synchrotron (DESY), Hamburg, Germany}

\begin{abstract}
The output SASE characteristics of the baseline European XFEL, recently used in the TDRs of scientific instruments and X-ray optics, have
been previously optimized assuming uniform undulators without considering the
potential of undulator tapering in the SASE regime. Here we demonstrate
that the performance of European XFEL sources can be significantly improved
without additional hardware. The procedure simply consists in the
optimization of the undulator gap configuration for each X-ray beamline.
Here we provide a comprehensive description of the soft X-ray photon beam
properties as a function of wavelength and bunch charge. Based on nominal
parameters for the electron beam, we demonstrate that undulator tapering allows
one to achieve up to a tenfold increase in peak power and photon spectral density in the conventional SASE regime. We illustrate this fact for the SASE3 beamline. The FEL code Genesis
has been extensively used for these studies. Based on these findings we suggest
that the requirements for the SASE3 instrument (SCS, SQS) and for the SASE3 beam transport system be updated.
\end{abstract}

%\begin{keyword}
%%edge radiation \sep near-field \sep electron-bunch diagnostics
%%\sep x-ray free-electron laser (XFEL)
%%\PACS 41.60.Cr \sep 42.25.-p \sep 41.75.-Ht
%\end{keyword}
%
\end{frontmatter}

\section{Introduction}

In this article we demonstrate that for nominal electron bunch distributions, the output radiation characteristics  of the European XFEL sources can be easily
improved compared to what has been assumed in the design reports of the scientific
instruments and the X-ray optics \cite{cdr-scs,cdr-sqs,cdr-optics}.  The output SASE characteristics of the baseline European XFEL have been previously optimized assuming uniform undulator settings.
However, in order to enable experiments over
a continuous photon energy range, European XFEL undulators are designed to be
tunable in photon energy by adjusting the gap \cite{xfel-tdr}. The availability of
very long tunable gap undulators at the European XFEL facility provides a unique
opportunity of up to tenfold increase in spectral density
and output power (up to the TW-level) for nominal electron beam parameter sets without modification to the baseline undulator design.

The technical note \cite{TSCH} provides an overview of the design considerations and the general layout of the X-ray instrumentation of the European XFEL sources, beam transport systems and instruments. 
Baseline parameters for the electron beam have been defined and presented in \cite{IG01}, \cite{feng}.
These parameters have been used for simulating FEL radiation characteristics and saturation lengths relevant to the European
XFEL SASE undulators \cite{YS1}. There the following definition of saturation was used:  "Saturation is reached at the magnetic length at which
the FEL radiation attains maximum brilliance. Beyond the saturation point,
the FEL operates in an over saturation mode where more energy can be
extracted from the electron beam at the expense of FEL parameters, including
bandwidth, coherence time, and degree of transverse coherence".

The approach based on the exploitation of the definition of the saturation point
reported above, and on the assumption that the best FEL parameters are reached at
saturation has been quite useful as a starting point for the analysis
of XFEL sources, beam transport and instruments.

One obvious way to enhance the SASE efficiency is to properly configure undulators with variable gap \cite{TAP1, TAP2, TAP3, TAP4}. In \cite{taper-xfel} it has been studied how a tapering procedure can be used to significantly
improve performance of the European XFEL sources without additional hardware.
The technique was demonstrated on  the example of the baseline SASE3 undulator considering  0.1 nC bunch and 2 keV photon energy. It was demonstrated that  undulator tapering allows one to achieve up to tenfold increase in peak power and photon spectral density at this particular nominal working point.

Tapering consists in a slow reduction of the field strength of the undulator  in order to
preserve the resonance wavelength, while the kinetic energy of the electrons
decreases due to FEL process. The undulator taper can be simply
implemented as discrete steps from one undulator segment to the next,
by changing the undulator gap.

The purpose of the present article is to give a more comprehensive description of
the SASE3 photon beam properties. We demonstrate that tapering allows one
to achieve up to a tenfold  increase in output at all achievable photon energies and all
nominal electron bunch charges.  A new set of baseline parameters of the
electron beam for the European XFEL has been recently updated \cite{IG01}, \cite{feng} . We present
a description of radiation properties generated by the SASE3 FEL undulator
driven by an electron beam with revised baseline parameters.
The SASE3 undulator has been placed behind SASE1 \cite{xfel-tdr}.
It is assumed that the electron beam is not disturbed by FEL interaction in the SASE1 undulator.
A method to control the FEL amplification process based on betatron switcher is described in \cite{beta-switch}.
Electron energies of 10.5 GeV, 14 GeV and 17.5 GeV have been assumed  for the calculation
of the baseline European XFEL operation. The lowest photon energies achievable in SASE3 are
then  250, 500, and 800 eV.  We highlight operation of SASE3 for the electron energy of 14 GeV.
\footnote{Recently, a modification of the main electron beam energy  operation points was made,
in order to account for the fact that the $K$ parameter of undulators at European XFEL turned out
to be systematically slightly smaller than designed. For simplicity, in this article we still consider
the old energy points. Since we highlight operation of SASE3 for the electron energy of 14 GeV, such change
will only affect the lower photon energy range in the calculations presented, which will now be achievable at $12$ GeV only.
Results, however will not change noticeably for the case study presented here, 
because the ratio between the two energies is small. For the same reason, 
the presented results can be applied to the 12 GeV working point with good accuracy. }

In the following we assume that SASE1 operates at the photon energy of 12 keV, and the FEL process is switched
off for dedicated SASE3 operation. Start-to-end
simulations \cite{IG01}, \cite{feng} give the electron beam parameters at the entrance of SASE1.
However, resistive wakefields up to the entrance of SASE3  modify the electron beam energy distribution and are therefore included in our simulations. Moreover,
as mentioned above, we assume that the lasing in SASE1 is inhibited with the help of the betatron switcher
technique, but the undulator gap is not opened. Therefore, energy spread due to quantum
fluctuations in SASE1 is accounted for as well.

We present a graphical overview of the main characteristics of SASE3 operating in the
tapered regime including  pulse energy, number of photons per pulse, peak power,
source size and divergence and maximum  value of photon spectral density as a function of
photon energy for different bunch charges.

The analysis of the nonlinear FEL process can be approached only numerically.
The Genesis code \cite{genesis} has been extensively used for our FEL studies. While the code has
been successful in reproducing results from LCLS experiments
and has been extensively benchmarked, next generation FEL codes like
ALICE \cite{alice} recently began to appear, taking advantage of more and
more advanced algorithms. In order to increase the confidence in  simulations
of the deep nonlinear regime in tapered undulators, it is instructive to
crosscheck Genesis results with ALICE. Despite the facts that simulation of deep
nonlinear SASE regime is a challenging problem for numerical analysis and that, moreover,
different codes adopt different numerical methods and random generators, it was
demonstrated (see \cite{taper-xfel}) that  differences in the output are well within the rms
($10 \%$) of shot-to-shot fluctuations.

Simulating the XFEL parameters for all working points is time-consuming. Apart from calculating the FEL process itself, several preliminary steps are to be taken: the electron optics in the undulators
is matched to yield optimal performance, undulator $K$ parameter settings are adjusted (including tapering), several other input parameters
such as mesh size are adjusted to achieve required precision. Moreover, standardization of input and output is important. It guarantees that a simulation procedure is reproducible, and could be further fed into
other calculation procedures. The mentioned issues  have been addressed by developing a python-based simulation framework \cite{agapov}.

To the users' benefit, it is important that updates in performance of various subsystems is quickly reflected onto the simulation results.
It is therefore foreseen that the complete up-to-date output for the SASE1 and SASE3 baseline
undulators for all electron beam energies will be maintained on the XFEL.EU photon beam parameter web page \cite{db}.

\section{SASE3 photon beam properties}

At the European XFEL facility three photon beamlines will be delivering X-ray pulses
to six experimental stations. The basic process adopted to generate the X-ray pulses is
SASE.  This section describes the source properties of the SASE3 undulator for the
soft X-ray beamlines at the European XFEL. The SASE3 undulator is 120 m
long and is expected to produce SASE FEL radiation  in the photon
energy range between  0.25 keV and 3 keV.

%The SASE3 undulator has been placed
%behind SASE1 \cite{TSCH}. It is assumed that electron beam is not disturbed by FEL interaction
%in SASE1 undulator. A method to control the FEL amplification process is based on
%betatron switcher described in [...]. Electron energies of 10.5 GeV, 14 GeV and 17.5 GeV
%have been assumed  for the calculation of baseline European XFEL operation [...].
%The lowest photon energies achievable in SASE3 are then  250, 500, and 800 eV.

As mentioned in the introduction we highlight operation of SASE3 for an electron energy of 14 GeV, which is the preferred
operation energy for the SASE1 beamline users. Since it will be necessary to run
the SASE1 and SASE3 beamlines at the same electron energy, this choice will
reduce the interference with SASE1 undulator line and increase the total
amount of scheduled beam time.

%The SASE3 characteristics for  different baseline
%electron beam energies will be available in [...], where we will present in more
%details an overview of radiation properties of all baseline (SASE1, SASE2, SASE3)
%radiators in tapered mode of operation driven by electron beam with baseline
%parameters.

A new set of baseline parameters of the electron beam for the European XFEL has
been updated recently. We present a description of radiation properties generated by
the SASE3 FEL undulator driven by an electron beam with the revised baseline
parameters presented in \cite{IG01}, \cite{feng}.  For fixed electron and photon energy, five working points are foreseen, corresponding to
bunch charges of $0.02$ nC , $0.1$ nC, $0.25$ nC, $0.5$ nC, $1$ nC, and resulting in pulse durations of roughly $2$ fs, $8$ fs, $20$ fs, $40$ fs and $80$ fs.

\begin{figure}[h!]
 \centering
 \includegraphics[width=120mm, height=90mm]{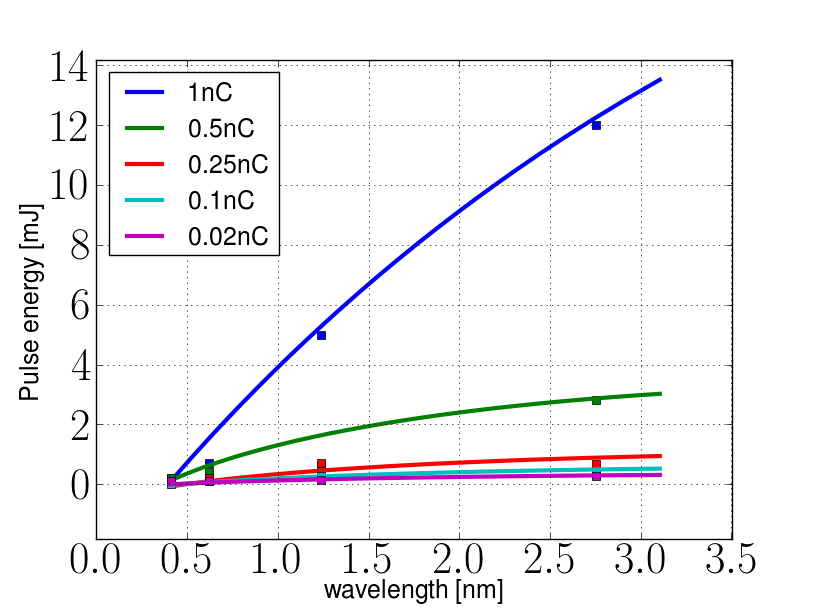}
 \includegraphics[width=120mm, height=90mm]{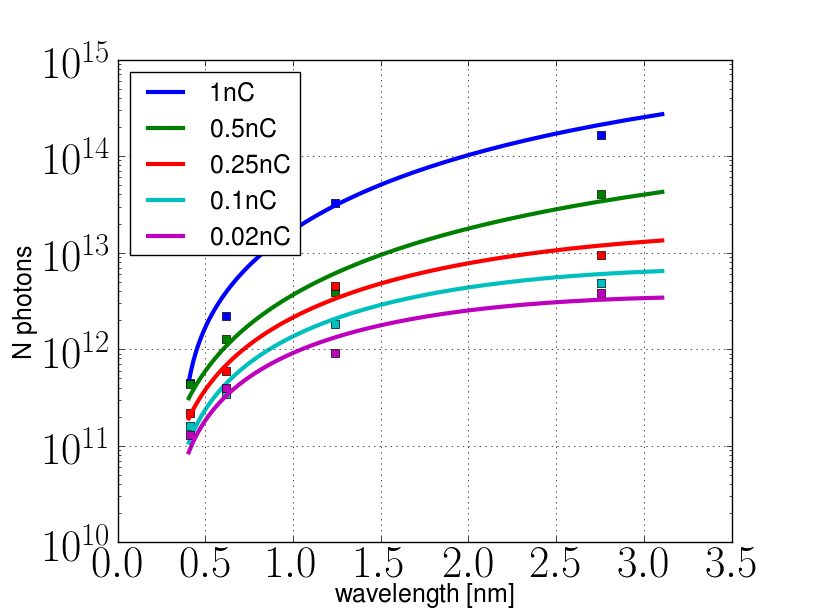}
 \caption{SASE3 baseline for 14 GeV electron energy:
 (top) pulse energy and (bottom) number of photons per pulse as a function of photon
 energy and bunch charge  in the SASE saturation mode of operation.}
 \label{fig:1}
\end{figure}

\begin{figure}[h!]
 \centering
 \includegraphics[width=120mm, height=90mm]{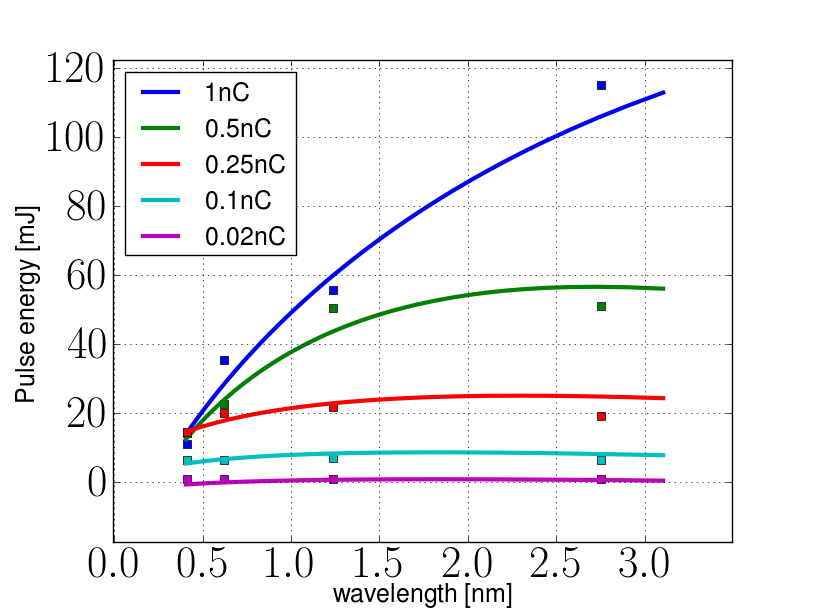}
 \includegraphics[width=120mm, height=90mm]{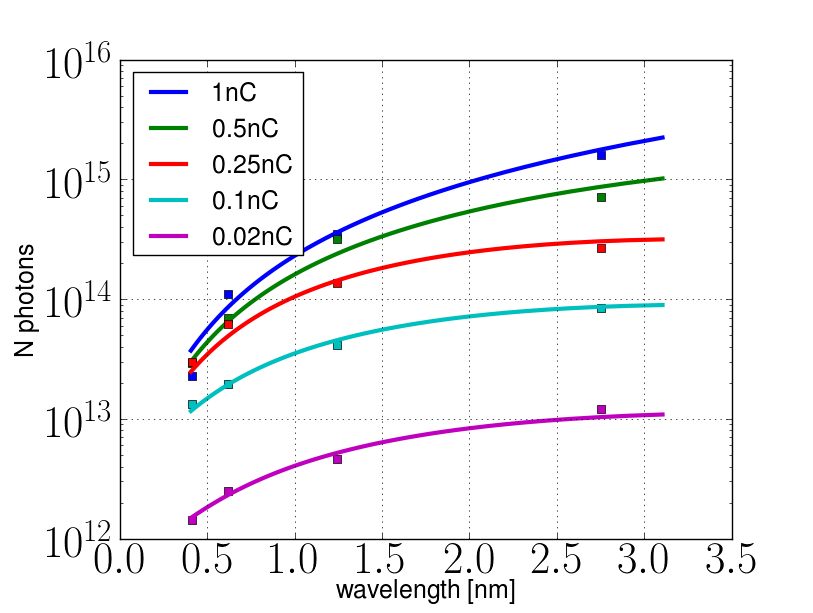}
 \caption{SASE3 baseline for 14 GeV electron energy:
 (top) pulse energy and (bottom) number of photons per pulse as a function of photon
 energy and bunch charge in the SASE tapering mode of operation.}
 \label{fig:2}
\end{figure}

The source properties: size, divergence, radiation pulse energy, and maximum
photon spectral density depend on photon energy, bunch charge, and electron energy.
The pulse energies and the number of photons per pulse are shown in Fig. \ref{fig:1}
for the tapered mode and in Fig. \ref{fig:2} for the saturation mode as functions of photon energy and bunch charge.
In the tapered mode, pulse energy (or, equivalently, number of photons) increases by up to ten times compared to saturation,
depending on the bunch charge and radiation wavelength. For short bunches (e.g. corresponding to $0.02$ nC)
the tapering efficiency drops since the radiation
slips forward relative to the electron bunch and stops being amplified.

%The pulse energy increases with bunch charge and decrease with photon energy.

\begin{figure}[h!]
 \centering
  \includegraphics[width=120mm, height=90mm]{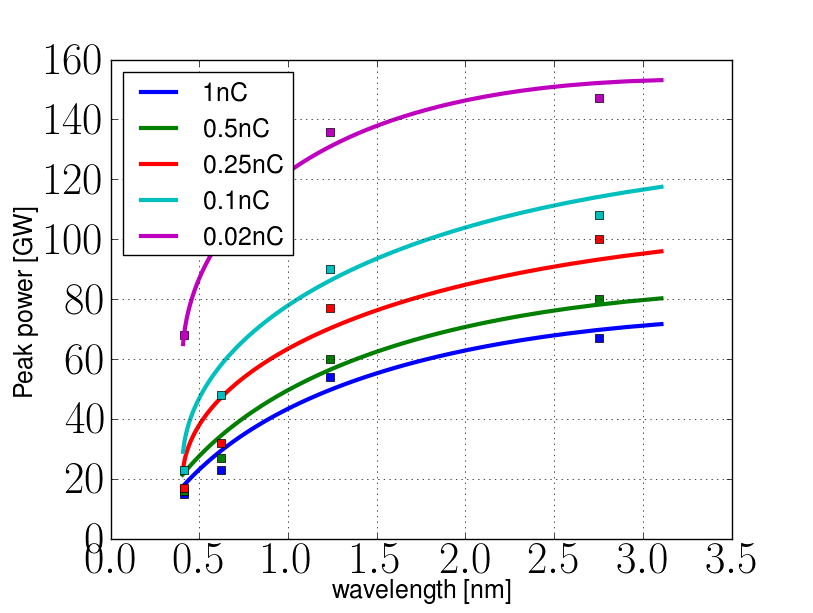}
  \includegraphics[width=120mm, height=90mm]{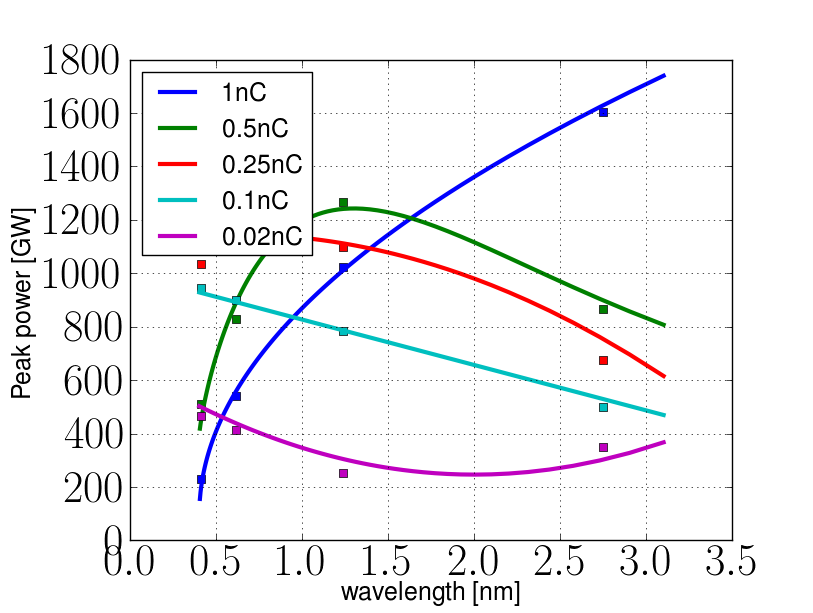}
 \caption{SASE3 baseline for 14 GeV electron energy: peak power in saturation (top) and tapering (bottom) mode. }
\label{fig:3}
\end{figure}

\begin{figure}[h!]
 \centering
  \includegraphics[width=120mm, height=90mm]{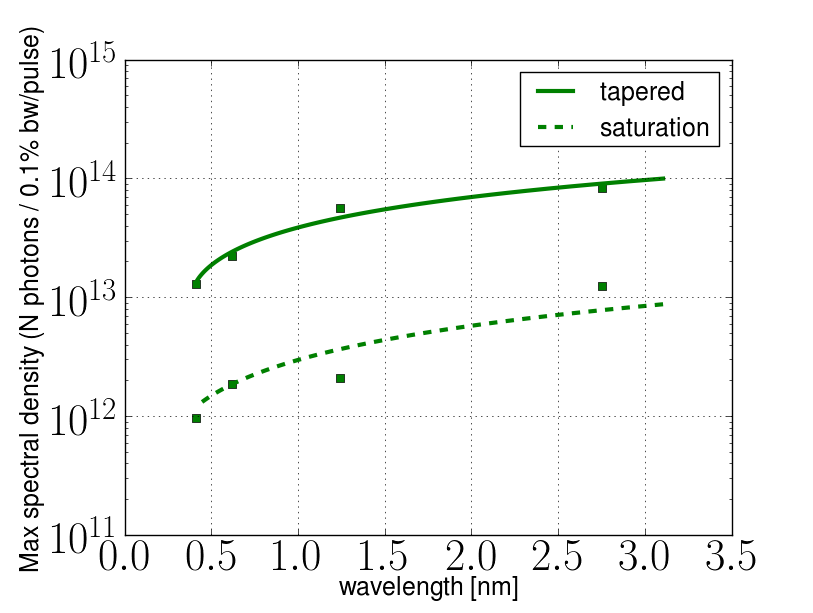}
 \caption{SASE3 baseline for 14 GeV electron energy: Maximum of average spectral density, for $0.5$ nC electron beam.}
\label{fig:4}
\end{figure}

Figs. \ref{fig:3} and \ref{fig:4}  show comparisons of peak power and photon spectral density produced
in the standard SASE mode at saturation and
in the tapered mode. Also in this case, up to tenfold increase in these parameters is observed.

For soft X-rays produced at SASE3 a grating monochromator will be used
in order to reduce the bandwidth of FEL radiation for spectroscopy applications.
This monochromator provides resolution better than $10^{-4}$ and is able to accept the high
power level of the XFEL radiation \cite{TSCH}. Since the monochromator line is much narrower than the SASE FEL line,
in order to predict the monochromator output in terms of number of photon per pulse  it is convenient to describe the
calculated spectral distribution by only one value, the maximum photon spectral density of the source.

\begin{figure}[h!]
 \centering
  \includegraphics[width=120mm, height=90mm]{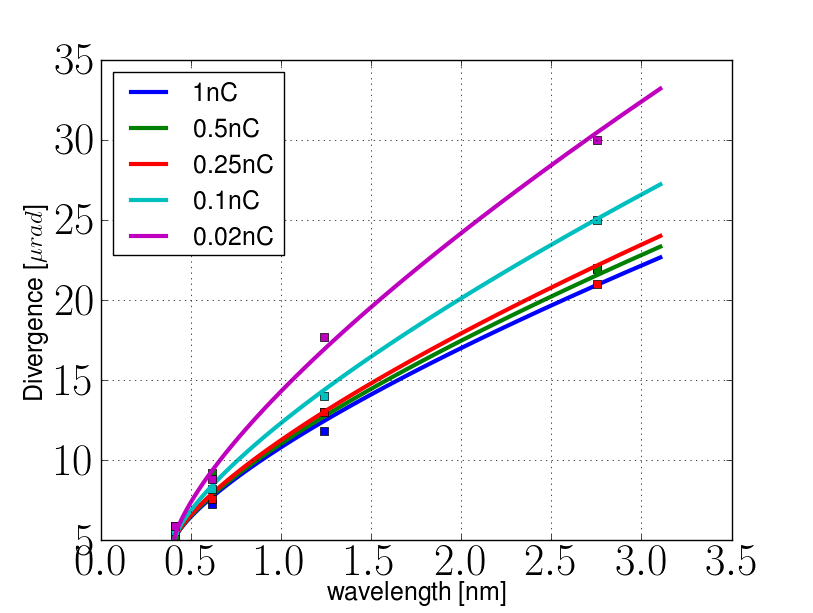}
  \includegraphics[width=120mm, height=90mm]{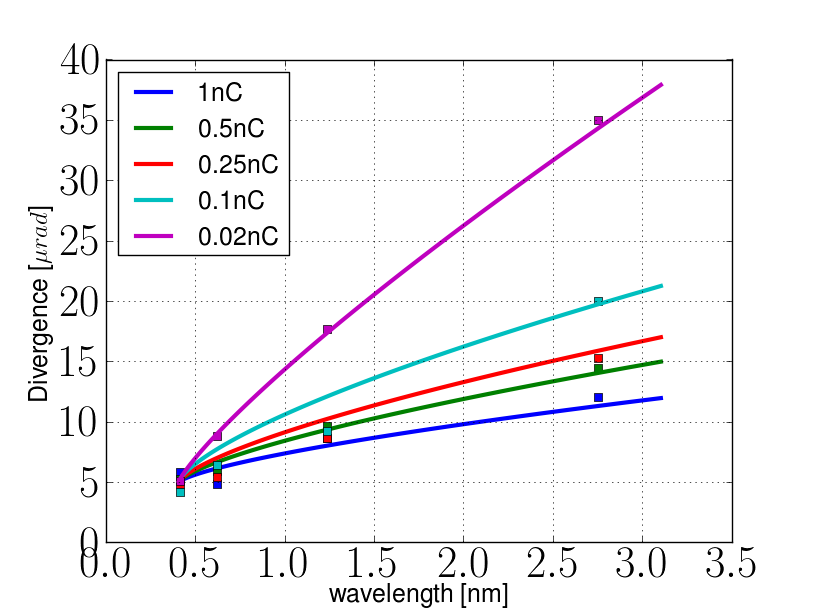}
 \caption{SASE3 baseline for 14 GeV electron energy: FWHM of angular distribution
of X-ray pulse energy as a function of photon energy and bunch charge in SASE saturation
mode (top) and in tapered mode (bottom).}
\label{fig:5}
\end{figure}

The source divergence is the most important parameter for the layout of the X-ray beam transport system. Fig. \ref{fig:5}
shows X-ray pulse divergence in terms of the FWHM of the angular distribution of X-ray pulse energy as a function of photon energy and bunch charge,
for saturation and tapered modes respectively. The source
divergence is largest for the smallest photon energies and the lowest bunch charges.
Since one needs to minimize diffraction from the optics aperture and preserve the
radiation wavefront, any optical elements should ideally have an aperture size large
enough to accept at least 4$\sigma$ tails.  The (horizontal) offset mirrors
of the SASE3 beamline are placed about 300 m behind the undulator exit. This mirror system
can be adjusted between 6 mrad and 20 mrad incidence angle. The X-ray optics and
transport group is planning to implement offset mirrors with clear aperture of 800 mm \cite{cdr-optics}.

With these parameters, using Fig. \ref{fig:5}, one obtains that the transverse clear aperture of the offset mirrors is in principle enough to fulfill
the 4$\sigma$ requirement for the SASE tapering mode of operation. The calculated radiation spot sizes at the undulator exit appear to be larger than
those in saturation mode. The exact size and profile make sense only in connection to studying focusing efficiency and
will be the subject of a separate study.

It is generally accepted that a variable beam size is the best approach
to make optimum use of the delivered photons per FEL pulse in each experiments.
The X-ray optical layout of the SASE3 instruments provides the option of
operating with a KB mirror focusing system. The phase distribution of the  FEL source
at the undulator exit is quasi-spherical. In this case, it is important to find the virtual FEL
source size and its position upstream the undulator exit corresponding to the maximum
pulse energy per unit surface in the source plane. If such virtual source will be placed in the
object plane of a focusing system one will reach in this case maximum energy density
in the image (sample) plane. Knowledge of virtual source size and its position is also
important for soft X-ray monochromator design. The calculated data allows to obtain the source position
using wavefront backpropagation, and this information will be added to the web pages \cite{db} later.

\section{FEL studies}

In this section we consider our simulation approach in more detail, providing illustrations for a particular working point.
As mentioned before, the nominal electron beam characteristics resulting from
start-to-end simulations in terms of current, emittance, energy spread and
energy can be found in \cite{IG01}, \cite{feng}. Additional energy chirp introduced by resistive
wakes in the SASE1 undulator vacuum chamber are included in our simulations,
as well as quantum diffusion effects in the SASE1 undulator. All simulations
were performed using the code Genesis \cite{genesis} running on a multiprocessor cluster. Results
are presented for the SASE3 FEL line, based on a statistical analysis. Large number of calculation points makes it
hard to perform statistically very accurate calculations even using high performance clusters.
Only few points were calculated with considerable statistics, which is typically 100 to 200 runs.
However, we believe than an accuracy of about $10-20\%$ is the best one can hope for with the present understanding of beam parameters.
Therefore, it turned out that for such characteristics as average number of photons and peak power, 20 runs are sufficient,
and most of the points were calculated with such statistics. Other parameters, e.g. pulse-to-pulse energy variation,
would require much more statistics and could be calculated separately for selected working points.

%consisting
%of 100 runs and comparing the SASE regime at saturation with the tapered SASE
%regime.

\begin{figure}[h!]
 \centering
 \includegraphics[width=60mm, height=40mm]{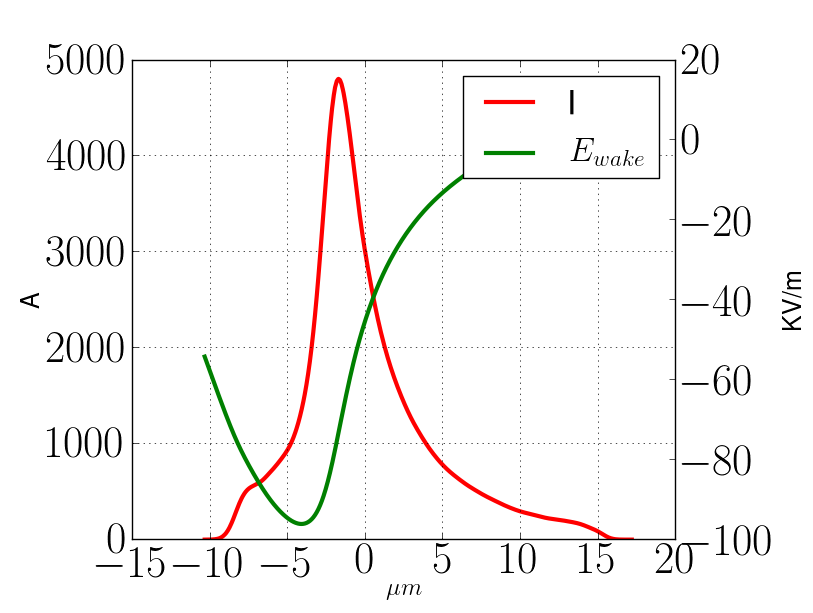}
 \includegraphics[width=60mm, height=40mm]{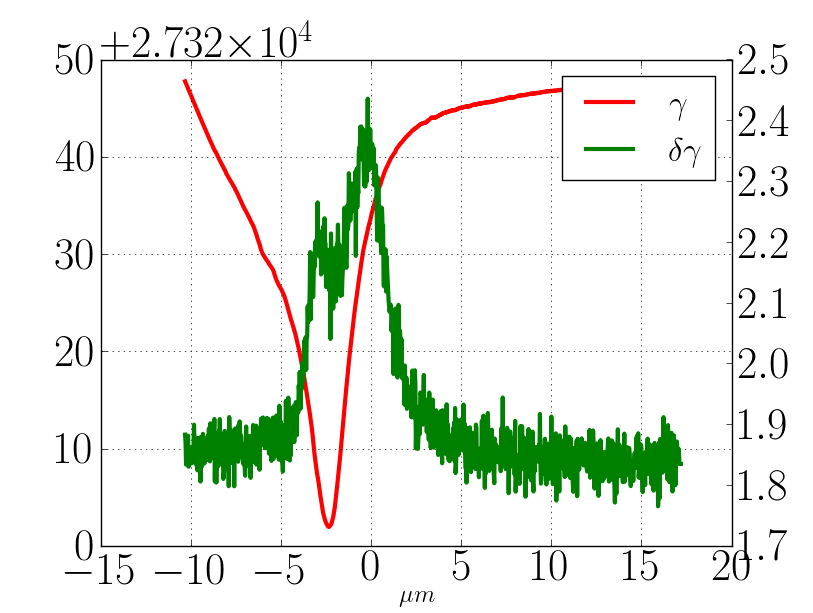}
 \includegraphics[width=60mm, height=40mm]{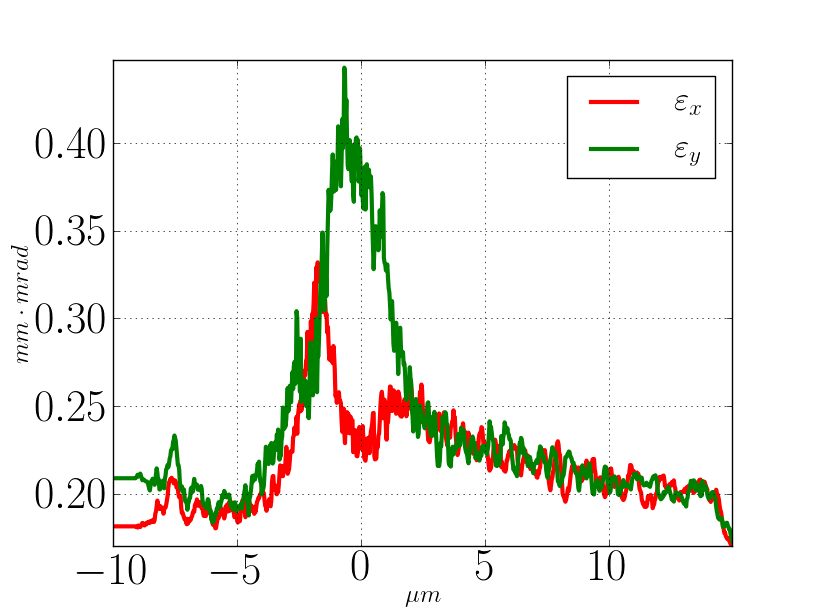}
 \includegraphics[width=60mm, height=40mm]{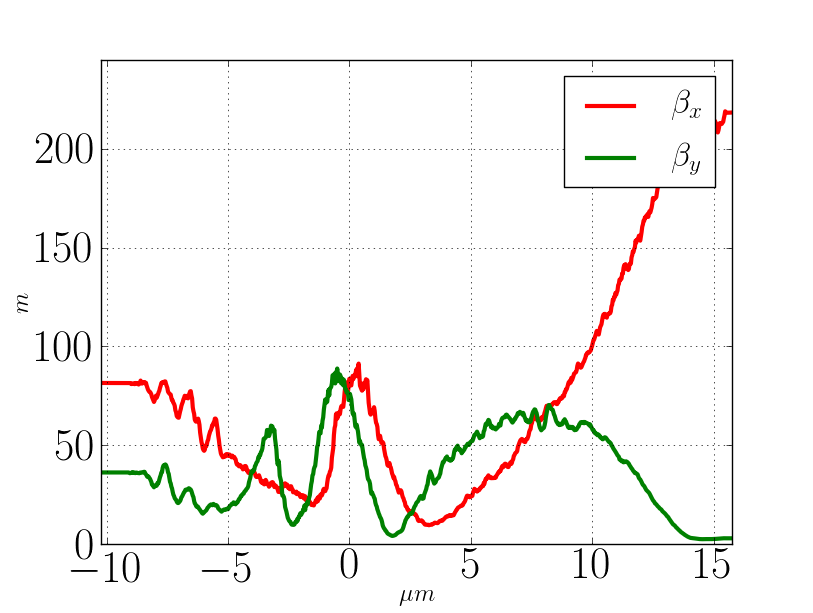}
 % screenshot.png: 2318x1277 pixel, 100dpi, 58.88x32.44 cm, bb=0 0 1669 919
 \caption{Simulated electron beam properties as functions of the position inside the bunch at the entrance of SASE3 (14 GeV, $0.1$ nC).
 Top left: current profile (red) and resistive wakefield (green).
 Top right: Energy (red) and energy variation (green) distribution.
 Bottom left: horizontal and emittances. Bottom right: horizontal and vertical $\beta$-functions.}
 \label{fig:7}
\end{figure}

Note that, according to beam dynamics simulation results,
the 6D phase space distribution of the electron bunch becomes very
involved and consequently the 3D electron bunch  is asymmetric. Because of that,
the output radiation pulse distributions are asymmetric too in both space-time
and reciprocal (angular-frequency) domains.  Previous numerical studies
of European XFEL photon beam characteristics \cite{YS1} were performed  for
Gaussian shape of electron peak current, uniform distribution of emittance
and energy spread and without including energy chirp, wakefields and undulator
tapering effects. According to this model of electron bunch and
undulator all FEL pulse distributions were symmetric with Gaussian-like shapes.
It will become important for users to be able to quantitatively characterize the
departure of a more realistic (i.e based on start-to-end- simulation results) pulse
distributions from Gaussian-like performance presented in  previous numerical
studies.

Here we illustrate in some depth the output distributions for the radiation pulse emitted at 2 keV
at nominal electron beam energy of 14 GeV, considering a 0.1 nC bunch.
The main electron and undulator parameters for simulations are shown in Table 1.
The nominal electron beam characteristics resulting from start-to-end
simulations are shown in Fig. \ref{fig:7} in terms of current, emittance, energy
spread and resistive wake in the SASE3 undulator.

\begin{figure}[h!]
 \centering
 \includegraphics[width=120mm, height=90mm]{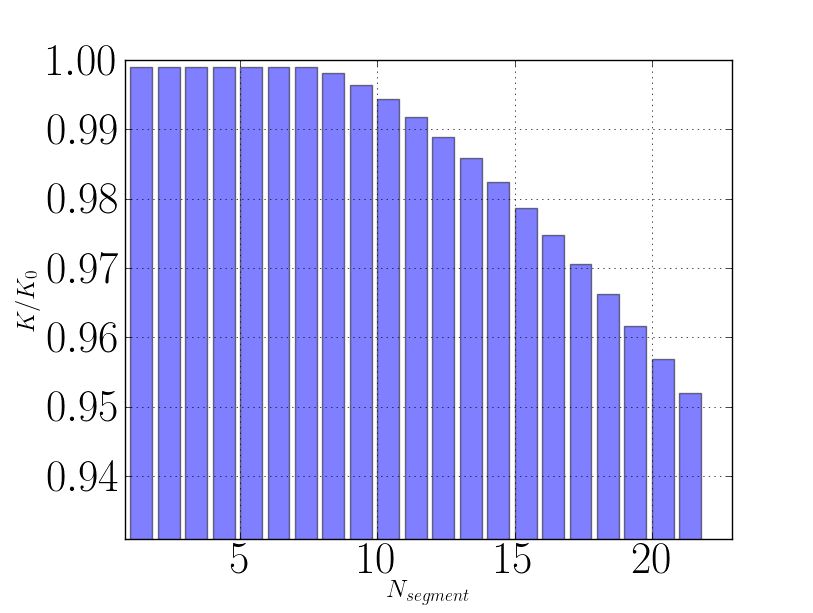}
 \caption{Example of a tapering function (SASE3 running at $14$ GeV, 0.1nC bunch charge, 1keV photon energy).}
 \label{fig:8}
\end{figure}

Tapering is implemented by changing the K parameter of the undulator segment by segment.
For each bunch charge and photon energy the tapering profile
is calculated separately, based on an optimization of theoutput power at the end of the undulator.
The tapering law used in this work has been found on an empirical basis. Two common possibilities are the power law

\begin{eqnarray}\nonumber
        K(n) &=& K_0, \quad   n < n_0\\
        K(n) &=&  K_1 + a_0 (n - n_0)^{a1},  \quad n \geq n_0 \nonumber
\end{eqnarray}

or the piecewise-quadratic law

\begin{eqnarray}\nonumber
K(n) =  K_{0i} + a_{0i} (n - n_i) + a_{1i} (n - n_i)^2, \quad n_{i} \leq n \leq n_{i+1}
\end{eqnarray}

where $n$ is the undulator segment count. For SASE3, the difference in using one or the other law is not significant and the piecewise-quadratic law was used for calculations.
The coefficients depend on the wavelength, but rather smoothly so the same setup is in principle effective over a wavelength range of about $500eV$.
A typical tapering function is presented in Fig. \ref{fig:8}.

\begin{figure}[h!]
 \centering
 \includegraphics[width=120mm, height=90mm]{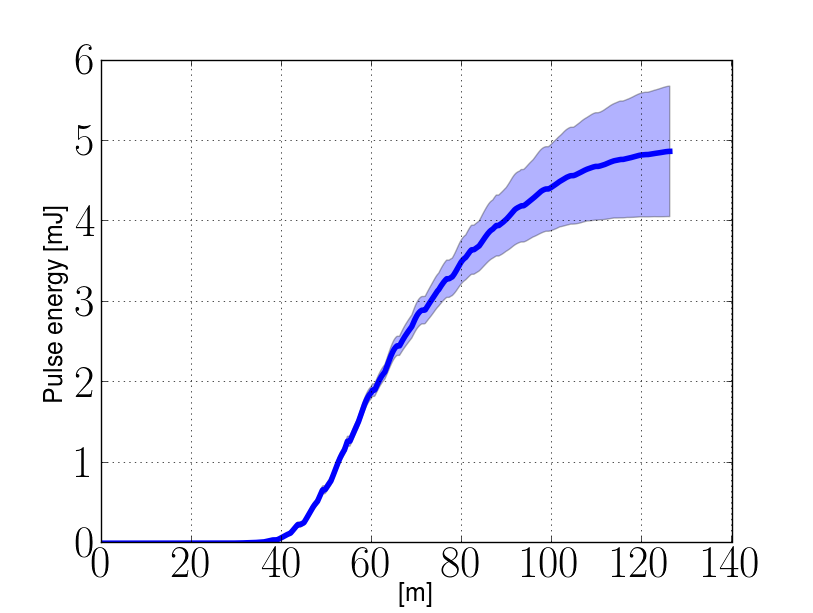}
 \caption{SASE3 baseline for 14 GeV electron energy, 0.1nC bunch charge, 1keV photon energy: Pulse energy evolution}
 \label{fig:10}
\end{figure}

\begin{figure}[h!]
 \centering
  \includegraphics[width=120mm, height=90mm]{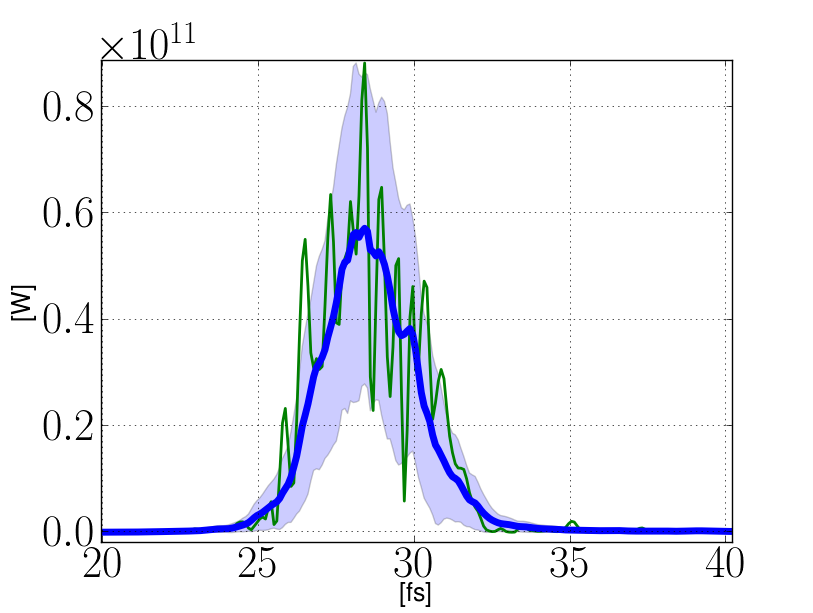}
  \includegraphics[width=120mm, height=90mm]{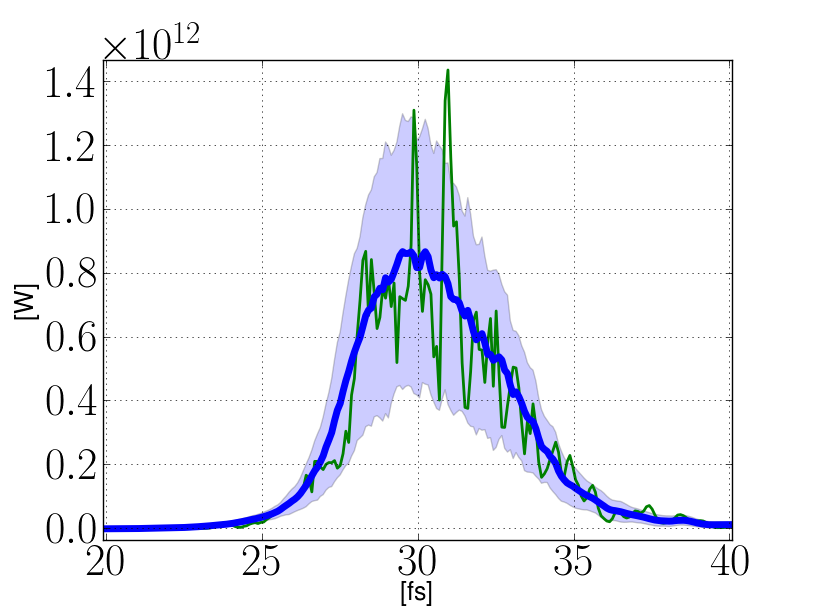}
 \caption{SASE3 baseline for 14 GeV electron energy, 0.1nC bunch charge, 1keV photon energy. Pulse shape: mean (blue), rms (shaded), and median (green). Saturation (top) and tapered (bottom).}
 \label{fig:11}
\end{figure}

\begin{figure}[h!]
 \centering
 \includegraphics[width=120mm, height=90mm]{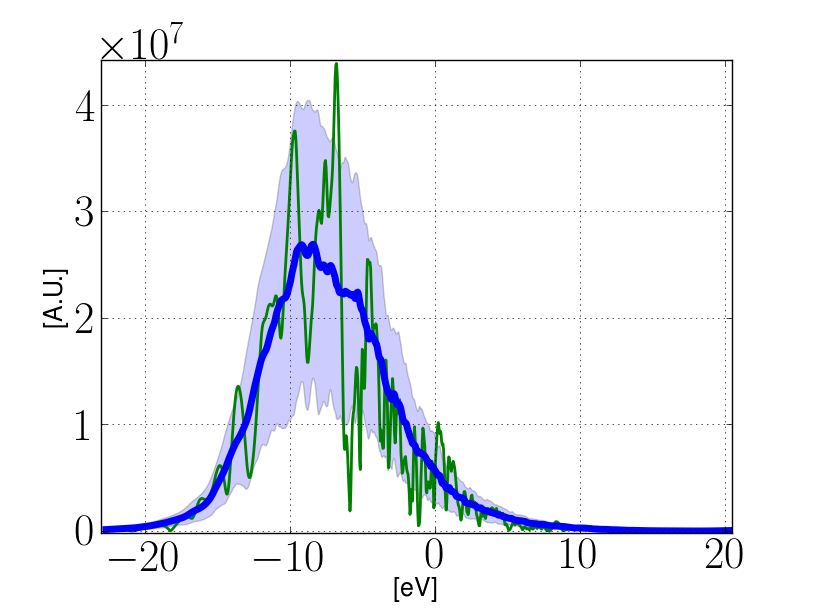}
 \includegraphics[width=120mm, height=90mm]{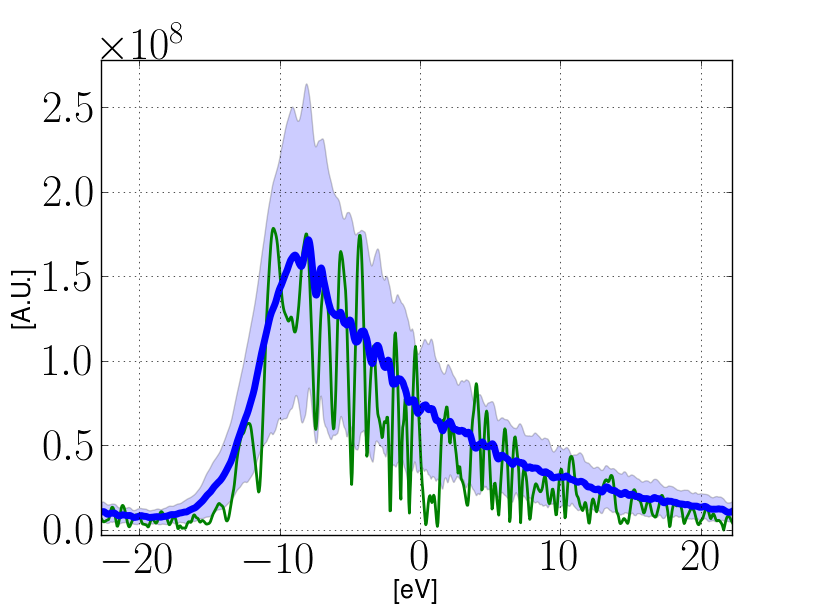}
 \caption{SASE3 baseline for 14 GeV electron energy, 0.1nC bunch charge, 1keV photon energy. Spectrum: mean (blue), rms (shaded), and median (green). Saturation (top) and tapered (bottom).}
 \label{fig:12}
\end{figure}

Figs. \ref{fig:10} shows the evolution of the output energy in the
photon pulse and of the variance of the energy fluctuation as a function of the
distance inside the  undulator, including tapering. Figs. \ref{fig:11} and \ref{fig:12} show a
comparison of power and spectrum produced in the standard SASE
mode at saturation (and, therefore, without tapering) and power and spectrum
produced in the SASE mode including post-saturation tapering. Note that up to tenfold increase in the shot-to-shot averaged spectral density
can be observed.

%Finally, in Fig. we show a comparison of radiation beam size (at the undulator exit)
%and divergence for the SASE operation mode in saturation, and for the SASE tapered
%mode.

% \begin{figure}[h!]
%  \centering
%  \includegraphics[width=120mm, height=90mm]{size_div_2kv_01nc.png}
%  \caption{Beam size (top) and divergence (bottom)}
%  \label{fig:13}
% \end{figure}
%
%
% \begin{figure}[h!]
%  \centering
%  \includegraphics[width=120mm, height=90mm]{05nc_1000ev_size.png}
%  \includegraphics[width=120mm, height=90mm]{05nc_1000ev_div.png}
%  \caption{Spectrum (mean and median). Saturation (top) and tapered (bottom)}
%  \label{fig:12}
% \end{figure}

\section{Conclusion}

XFELs are relatively complicated devices and their description is involved. The evolution of different approaches
to software for design and analysis of particle accelerators, beam
transport system, and FEL sources suggests that publicly available packages
should be chosen rather than proprietary software. Source
code and user's manual should be available. There are many reasons
for the codes to be publicly available. This guarantees that the conditions
of simulation studies will be publicly known and reproducible. Additionally, in this way many
improvements to the codes have resulted through international collaborations
with users and through regular feedback regarding how the codes are serving
the needs of the accelerator community.

In this article we presented the results of the initial groundwork for the standardization
of the FEL code to facilitate further international collaborations and comparisons in
the simulation results. The publicly available package Genesis \cite{genesis} has been chosen
for European XFEL sources simulations. The continued high-level use of the above
mentioned code demonstrates its great value to the international
XFEL community. This code has been experimentally verified by working hardware
at SLAC and other laboratories around the world. Special efforts were made
towards standardizing the input and output format for Genesis and various beam physics codes.
This  also increases the flexibility of the code application, and opens the possibilitiy to directly use electron beam characteristics
from start-to-end simulations. Continuing towards code standardization would be of great benefit
at the European XFEL, especially on the stage of facility commissioning.

In this article we demonstrated that the potential of the European XFEL in the
standard SASE mode has been underestimated up to the present day. In other words,
the output X-ray pulse parameters indicated in the design reports of scientific
instruments (SQS, SCS) and X-ray beam transport system are far from the optimum found in this
paper. Based on start-to-end simulations it has been shown that tapering of baseline SASE3
undulator provides an additional factor of ten increase in spectral density and output
power (up to TW-level) for a baseline electron beam parameter set.

%\paragraph{Statistics}

%Total radiation power distribution

%\begin{equation}
%p(W) =  \frac{M^M}{\Gamma(M)} \left( \frac{W}{\left< W \right>} \right)^{M-1} \frac{1}{\left< W \right>} \exp(-M \frac{W}{\left< W \right>})
%\end{equation}

%number of 'longitudinal modes'

%\begin{equation}
%\frac{1}{\sqrt{M}} = \frac{\Delta \omega }{\omega} \approx 2 \rho
%\end{equation}

% \paragraph{Uncertainties}
%
% Uncertaities/jitters in beam and undulator parameters are hard to account for before the machine is. It seems reasonable to assume that these will be
% understood and diminished in the process of machine operations, but that experimentalists should be aware on the errorbars on the predicted parameters.
% So, for example, in figure x cumulative effect of $e 10^{-4}$ variation in $K$ and $10^{-4}$ variation in beam energy is presented, based on steady
% state simulations. The effect of emittance and optics uncertainty can be roughly estimated from Figures 1,2.
%
% \begin{figure}[h!]
%  \centering
%  \includegraphics[width=90mm, height=60mm]{sase3_error_scatter.png}
%  % screenshot.png: 2318x1277 pixel, 100dpi, 58.88x32.44 cm, bb=0 0 1669 919
%  \caption{Cumulative effect of energy jitter and K variation from table 1, 200 Monte-Carlo realizations}
%  \label{fig:0}
% \end{figure}

\end{document}